\documentclass{aastex}        
\usepackage{spr-astr-addons}

\begin{document}
\title{Shadow of Kerr-Taub-NUT black hole}

\shorttitle{Shadow of black hole} \shortauthors{Abdujabbarov et
al.}

\author{Ahmadjon Abdujabbarov\altaffilmark{1,2}} \email{ahmadjon@astrin.uz}  \and
\author{Farruh Atamurotov\altaffilmark{1,3}}   \email{fara_90-91@mail.ru} \and
\author{Yusuf Kucukakca \altaffilmark{4}} \email{ykucukakca@gmail.com}
\author{Bobomurat Ahmedov\altaffilmark{1,2}} \email{ahmedov@astrin.uz}
\author{Ugur Camci\altaffilmark{4}} \email{ucamci@akdeniz.edu.tr}

\altaffiltext{1}{Institute of Nuclear Physics,
        Ulughbek, Tashkent 100214, Uzbekistan}
\altaffiltext{2}{Ulugh Begh Astronomical Institute,
Astronomicheskaya 33, Tashkent 100052, Uzbekistan}
\altaffiltext{3}{National University of Uzbekistan,  Tashkent
100174, Uzbekistan} \altaffiltext{4}{Department of Physics,
Akdeniz University, 07058 Antalya, Turkey}

\begin{abstract}

The shadow of a rotating black hole with nonvanishing
gravitomagnetic charge has been studied. It was shown that in
addition to the angular momentum of black hole the gravitomagnetic
charge term deforms the shape of the black hole shadow. From the
numerical results we have obtained that for a given value of the
rotation parameter, the presence of a gravitomagnetic charge
enlarges the shadow and reduces its deformation with respect to
the spacetime without gravitomagnetic charge. Finally we have
studied the capture cross section for massive particles by black
hole with the nonvanishing gravitomagnetic charge.

\end{abstract}


\section{Introduction}\label{intro}

Until now it is not found any observational proof of existence of 
gravitomagnetic monopole, i.e. so-called NUT
\citep{nut63} charge. 
Investigation of the massive and massless particles motion in NUT 
spacetime may provide  tool for studying new
important general relativistic effects which are associated with
nondiagonal components of the metric tensor and have no Newtonian
analogues. \citet{zonoz07,kkl08,ma08} studied the solutions for
electromagnetic waves and interferometry in spacetime with NUT
parameter. \citet{aliev08} considered Kerr-Taub-NUT spacetime with
Maxwell and dilation fields. 
In our preceding papers~\citep{mak08,aak08} we have studied the
plasma magnetosphere around a rotating, magnetized neutron star
and charged particle motion around compact objects immersed in
external magnetic field in the presence of the NUT parameter. 
The Penrose process in the spacetime of rotating black hole with
nonvanishing gravitomagnetic charge has been considered
by~\citet{shaymiev11}.

The geodesics 
of test 
charged (both electrically and magnetically charged) test particles in the Reissner-Nordstr\"{o}m and Taub-NUT
space-times fully analyzed by~\citet{lera1,lera2,claus}.
\citet{frolov1}  considered higher-dimensional black hole spacetimes and
null geodesics. 
In refs.~\citep{frolov2,frolov3,frolov4} the
parallel transport equations in the higher-dimensional
Kerr-NUT-(A)dS spacetimes have been studied in detail. 
Magnetized black hole on the
Taub-NUT instanton has been considered by~\citet{nedkova}. 
\citet{virmani} made detailed analyse of few geometrical properties of Taub-NUT
space-time metric.
\citet{amarilla10,amarilla12,amarilla13} have studied the shadow of rotating black holes in Chern-Simons modified gravity, braneworld gravity and Kaluza-Klein model.

In this paper 
we consider 
photon motion and circular orbits around the rotating
NUT black
hole and its shadow paying attention to the influence by gravitomagnetic charge. 
The paper
organized as follows: 
first, in Sect. 2 we analyse the geodesics of massless 
particles in the Kerr-Taub-NUT space-time metric. 
Sect. 3 is devoted to study the shapes of the shadows of rotating 
black holes for the different
values of the rotating parameter and gravitomagnetic charge. 
In Sect. 4 we study the capture cross section for massive
particles by black hole with nonvanishing gravitomagnetic charge.
We conclude and discuss all obtained results of the paper~in Sect. 5.
In this paper a space-like signature $(-,+,+,+)$ and a system of
units in which $G = 1 = c$ have been used. 
Greek indices are taken to run from 0 to 3 
and
Latin
indices from 1 to 3; 
covariant derivatives are denoted with a
semi-colon and partial derivatives with a comma.

\section{Photon motion around rotating Taub-NUT black holes}

In this section we will study 
massless particles  
motion around 
a rotating black hole with the total mass 
$M$ 
in the presence of non vanishing 
gravitomagnetic charge. 
This black hole 
is described by the space-time metric~\citep{nut63,mak08,aak08}:
\begin{eqnarray}
ds^2 & 
=
&  -\frac{1}{\Sigma} \left(\Delta - a^2 \sin^2\theta
\right)dt^2 + \Sigma \left(\frac{dr^2}{\Delta} +
d\theta^{2}\right)\nonumber \\
&& 
+ \frac{1}{\Sigma}\left[ (\Sigma
+a\chi)^2 \sin^2\theta  -
\chi^2\Delta \right]d\phi^2
\nonumber \\
&& +\frac{2}{\Sigma}(\Delta\chi
-a(\Sigma+a\chi)\sin^2\theta)dt
d\phi. \label{metric}
\end{eqnarray}
In above expression the following notations $\Delta, \Sigma,$ and $\chi$ are defined as
\begin{eqnarray}
 \Delta &=& r^2 + 
a^2  - 
l^2
- 2 M r , \nonumber \\ \nonumber \\
 \Sigma&=&
r^2+
(l+a \cos\theta)^2 , \nonumber \\ \nonumber \\
 \chi&=&
a \sin^2\theta- 2 l \cos\theta,   \nonumber
\end{eqnarray}
and $ a $ is the specific angular momentum of the black hole ($ a=J/M $) and $ l $ is the gravitomagnetic charge.
One 
can determine the event horizon by largest root of the  
equation
$\Delta=0$. The solution has the following form 
\begin{equation}
r_{+}= M + (M^2 - a^2+l^2)^{1/2}. \label{horizon}
\end{equation}

\begin{figure}[htb]
\begin{center}
\includegraphics[width=8cm]{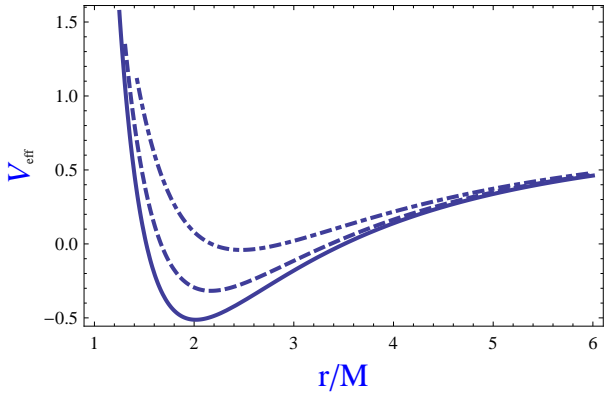}
\caption{The radial dependence of the effective potential of
radial motion of the massless particles for the different values
of the gravitomagnetic charge: solid line for $l/M=0.1$, dashed
line for $l/M=0.5$, and dot-dashed line for $l=0.9$.\label{rotr}}
\end{center}
\end{figure}

It is obvious that light coming from the distant source will be deflected by the gravitational influence by the black hole originated between observer and the source of the light. The deflection is going to increase with the decreasing of impact parameter of the photons and eventually the photon emitted by the distant source can be captured by black hole. At the end this effect cause no-light zone or dark zone in the sky due to the existence of black hole between observer and light source. The above mentioned dark zone is called shadow of the black hole and the shape of it is totally defined geodesics of massless particles. one can use the  Hamilton-Jacobi equation to obtain the equation of motion of photons in the given space-time metric, in our case in space-time metric of rotating black hole with non vanishing gravitomagnetic charge:
\begin{eqnarray}
\label{hamjam} 
\frac{\partial S}{\partial
\tau}
&=&
-
\frac{1}{2}
g^{\alpha\beta}\frac{\partial S}{\partial
x^{\alpha}}\frac{\partial S}{\partial x^{\beta}},
\end{eqnarray}
here $\tau$ is an affine parameter along the null-geodesics.
In
order to separate variables (the separable problem of
Hamilton-Jacobi equation in Kerr-Taub-NUT 
spacetime has been
studied by~\citet{dt02}), one can choose the 
action for photons in the following form:
\begin{eqnarray}
\label{acthamjam}
S &=& 
\frac{1}{2}
m^2 \tau - 
{\cal E} t + 
{\cal L} \phi +
S_{\rm r}(r)+S_{\rm \theta}(\theta) \ ,
\end{eqnarray}
here $m$ is the mass of a test particle. 
By ${\cal E}$ and ${\cal
L}$ we designated the 
energy and the angular momentum of 
the particle, respectively.
Putting the rest mass of photon as 
zero $m =0$ one may  solve the  
Hamilton-Jacobi equation for null-geodesics
\begin{eqnarray}
\Sigma\frac{dt}{d\tau}
&=&
\frac{r^2+a^2+l^2}{\Delta}\left[(r^2+a^2+l^2){\cal E} -a {\cal L}
\right] \nonumber\\
&& + \frac{\chi}{\sin^2\theta}({\cal L} -\chi {\cal E} ) ,
\label{reteng}
\\
\Sigma\frac{d\phi}{d\tau}&
=
&\frac{a}{\Delta}\left[(r^2+a^2+l^2){\cal
E}
-a {\cal L} \right] \nonumber\\
&&+ \frac{1}{\sin^2\theta}({\cal L} -\chi {\cal E} ), \label{refii}
\\
\Sigma\frac{dr}{d\tau}&=
&
\sqrt{\mathcal{R}}, \label{rerr}
\\
\Sigma\frac{d\theta}{d\tau}
&=&
\sqrt{\Theta}, 
\label{etete}
\end{eqnarray}
here $\mathcal{R}(r)$ and $\Theta(\theta)$ are introduced
notations and they have the following form:
\begin{eqnarray}
\mathcal{R}
&=&
\left[(r^2+a^2+l^2){\cal E} -a {\cal L}
\right]^2-\Delta\left[\mathcal{K}+({\cal L} -a {\cal E})^2\right],
\
\\
\Theta
 &=&
 \mathcal{K}+\cos^2\theta\left [(a^2-\frac{4
l^2}{\sin^2\theta}){\cal E}^2-\frac{{\cal
L}^2}{\sin^2\theta}\right] \nonumber\\
&&
+4 l {\cal E} \cos\theta\left({\cal E} a
-\frac{{\cal
L}}{\sin^2\theta}\right) ,
\end{eqnarray}
where  $\mathcal{K} $ is Carter constant.
Defining the effective potential for massless particles as
$({dr}/{d\tau})^2 = V_{\rm eff}$ one may study the radial
motion of photons in the presence of gravitomagnetic charge. In
the Fig. \ref{rotr} the radial dependence of the effective
potential of radial photon motion is shown. From the figure it is
seen that with the incease of  the gravitomagnetic charge the
shape of the effective potential is going to shift to the observer
at infinity. This corresponds to increasing the event horizon of
the Kerr-Taub-NUT black hole. Moreover, one may conclude from the
Fig. \ref{rotr} that with the increase of the gravitomagnetic
charge the circular photon orbits become unstable.

Photon motion around rotating black hole with 
non vanishing gravitomagnetic charge can 
be described using the expressions~(\ref{reteng})--(\ref{etete}).
One may easily introduce the following
two impact parameters:  $\xi={\cal L}/{\cal E}$,
$\eta=\mathcal{K}/{\cal E}^{2}$ in 
order to clarify the photon motion in complete way. 
Now it useful to use the equation (\ref{rerr}) for defining the 
shape of the  dark zone created by rotating black hole with 
non vanishing gravitomagnetic charge. 
The
condition of being boundary of shadow is the following: 
$\mathcal{R}(r)=0=d\mathcal{R}(r)/dr$. 
Using this condition one can obtain the following equations:
\begin{eqnarray}
\xi(r)
&=&
\frac{a^{2}(1+r)+r^{2}(r-3)+l^{2}(1-3 r) }{a(1-r)},
\label{eqxi}
\\
\eta(r)
&=&
\frac{1}{a^{2}(r-1)}\bigg\{r^{3}[4a^{2}-r(r-3)^{2}]-l^{2}\big[4
r^2 a^2\nonumber
\\
&+
&(1-3 r)(l^2(1-3 r)-6 r^2+4 a^2 r+2 r^3)\big]\bigg\} \ . \ \ \ \
\label{eqetata}
\end{eqnarray}

\section{Kerr-Taub-NUT black hole shadow}

\begin{figure}[htb]
\begin{center}
\includegraphics[width=8cm]{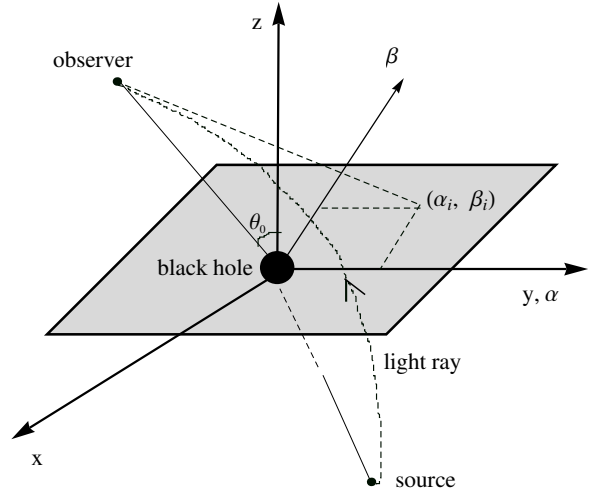}
\caption{The scheme of the gravitational lens system. A reference
coordinate system $(x, y, z)$ with the black hole at the origin can be set up by an
observer far away from the black hole. The straight continuation of observing light ray intersects the plane $\alpha-\beta$  at the
position $(\alpha_{i}, \beta_{i})$. \label{coordinates}}
\end{center}
\end{figure}

\begin{figure*}
\begin{center}
a. \includegraphics[width=8cm]{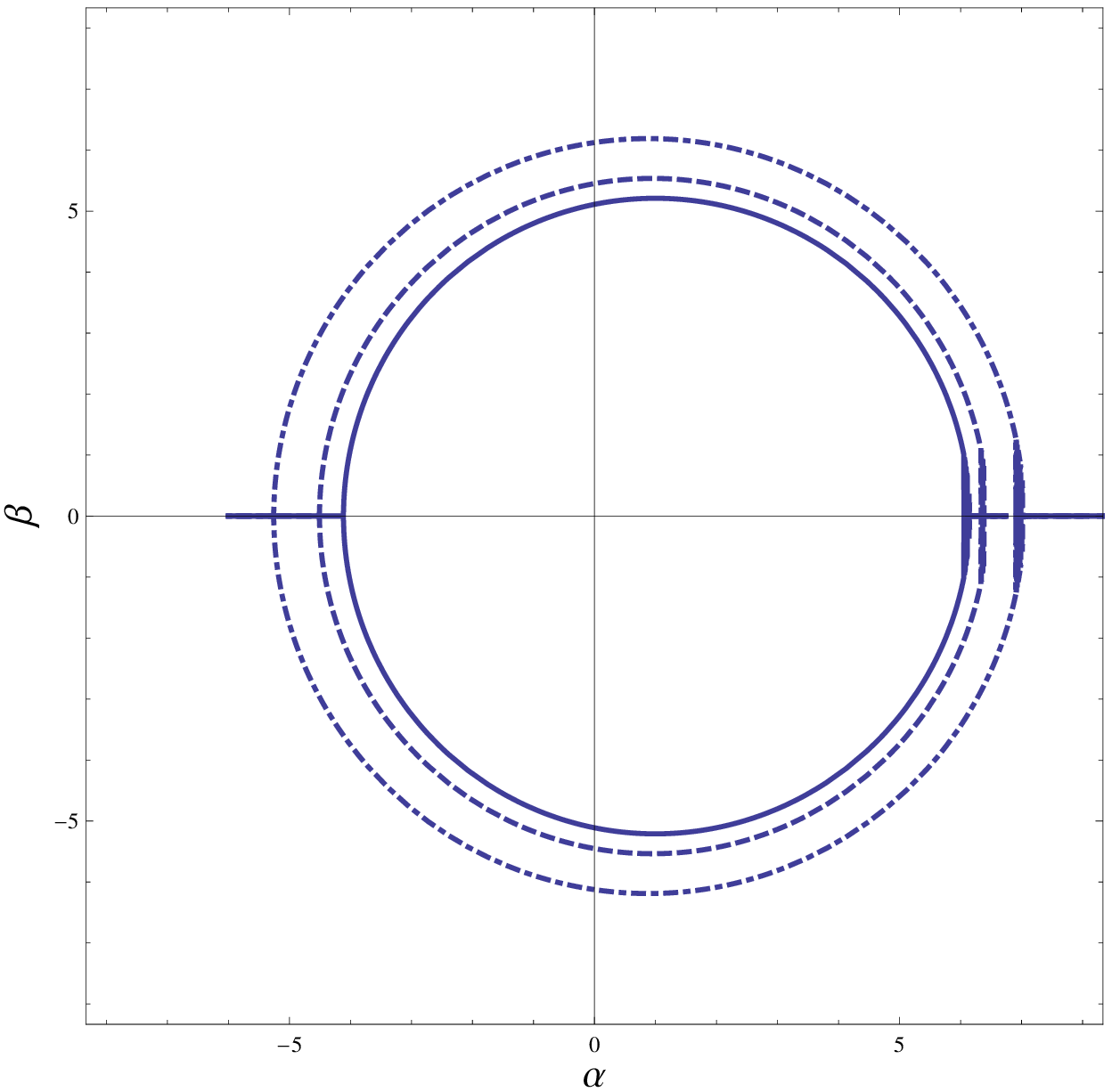}
b. \includegraphics[width=8cm]{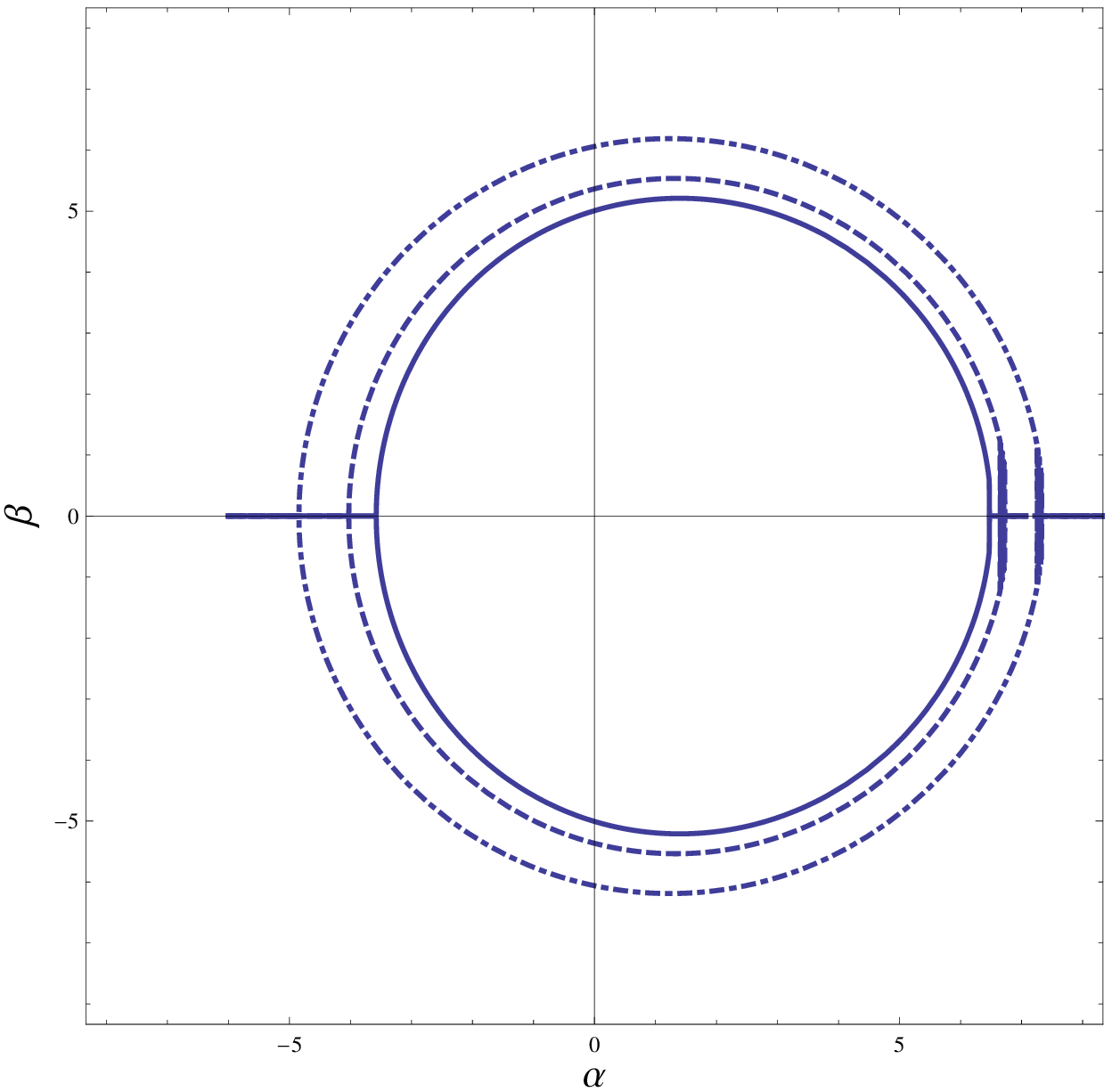}

c. \includegraphics[width=8cm]{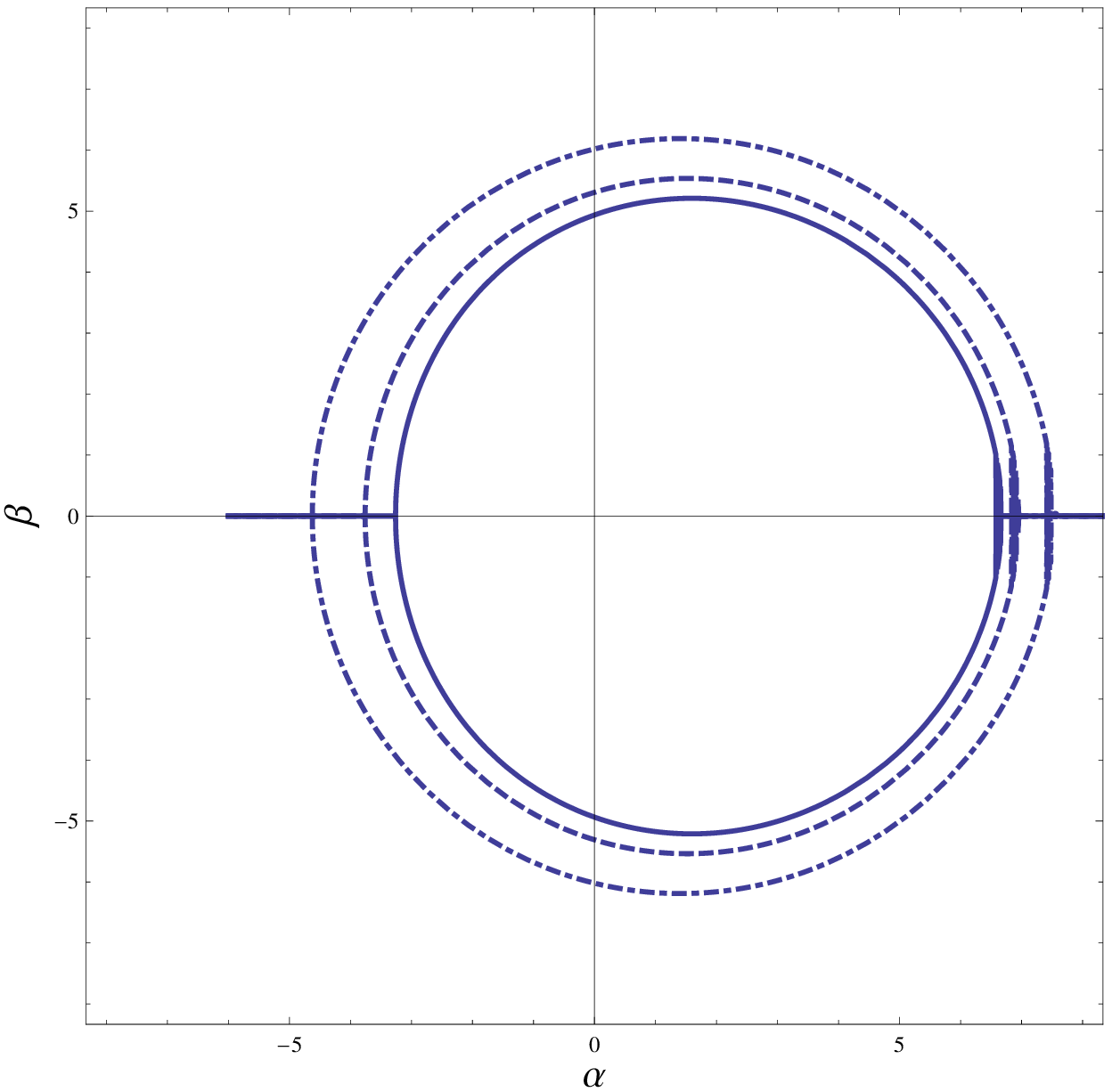}
d. \includegraphics[width=8cm]{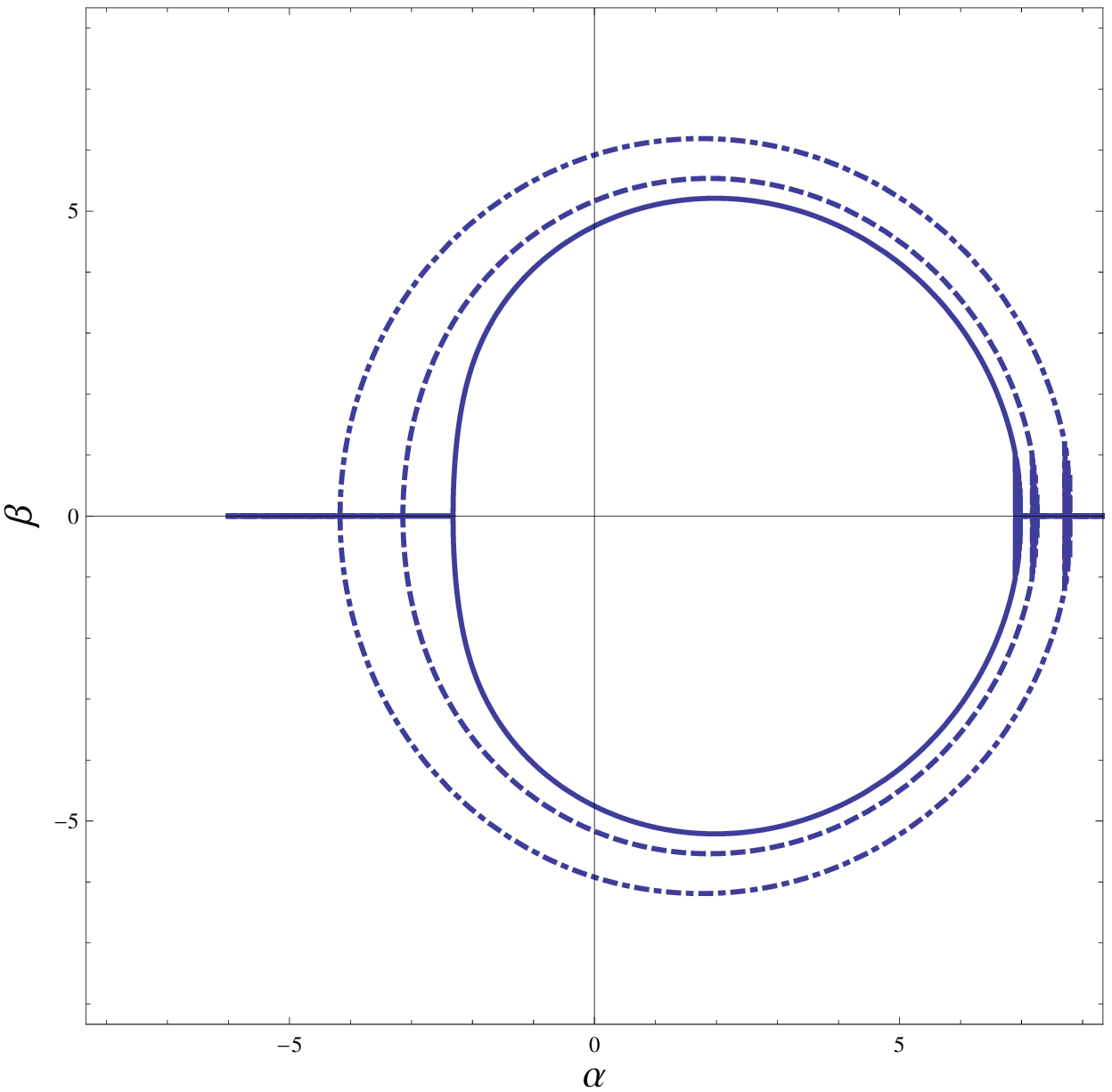}

\caption{The shapes 
of the rotating black hole 
shadow with non vanishing 
NUT charge  originated 
at centre of coordinates
when
$\theta=\pi/2$. 
(a): $a/M = 0.5$, 
(b):  $a/M = 0.7$,  
(c):  $a/M = 0.8$, and  
(d): $a/M
= 0.99$. 
In all figures  
solid lines correspond to $l/M=0.1$, 
dashed lines correspond to $l/M=0.5$ , and
dashed-dotted  lines correspond to $l/M=0.9$. 
The region bounded by each
curve corresponds to the balk hole shadow. \label{aylanalar}}
\end{center}
\end{figure*}

To obtain the silhouette of the rotating black hole with non vanishing NUT charge it is very convenient to use the celestial coordinates~(\cite{vazquez}):
\begin{eqnarray} 
 \label{alpha}
\alpha&=& \lim_{r\rightarrow \infty}\left(
-r^{2}\sin\theta\frac{d\phi}{dr}\right)\ ,\\
{\rm and}&& \nonumber \\
 \label{beta1}
\beta
&=&\lim_{r_{0}\rightarrow \infty}r_{0}^{2}\frac{d\theta}{dr}\ ,
\end{eqnarray}
if the distant observer
is located at long 
distance from
the rotating black
hole with NUT charge
then limit  $r\rightarrow\infty$ can be 
used~\citep{amarilla10,amarilla12,amarilla13}. 
The geometrical structure of celestial 
coordinates is schematically shown in 
Fig.~\ref{coordinates}. 

Calculating $d\phi/dr$ and 
$d\theta/dr$ using the spacetime
metric (\ref{metric}) and putting the 
results into (\ref{alpha}),
(\ref{beta1}), using (\ref{refii}), (\ref{rerr}), 
and (\ref{etete}) 
we obtain equations 
for $(\alpha, \  \beta)$ 
coordinates 
in the following form:
\begin{eqnarray}  \label{alphapsi1}
\alpha&=&-\xi\csc\theta_{0}\ ,
\\
{\rm and}&&\nonumber\\
 \label{beta2}
\beta&=&\pm \bigg[\eta +(a^2-\frac{4 l^2}{\sin^2\theta_{0}})\cos
^{2}\theta_{0}-\xi^{2}\cot ^{2} \theta_{0}\nonumber\\
&&+4 l \cos\theta_{0}( a-\frac{\xi}{\sin^2\theta_{0}})
\bigg]^{1/2}\ ,
\end{eqnarray}
%

\begin{figure}[htb]
\begin{center}
\includegraphics[width=8cm]{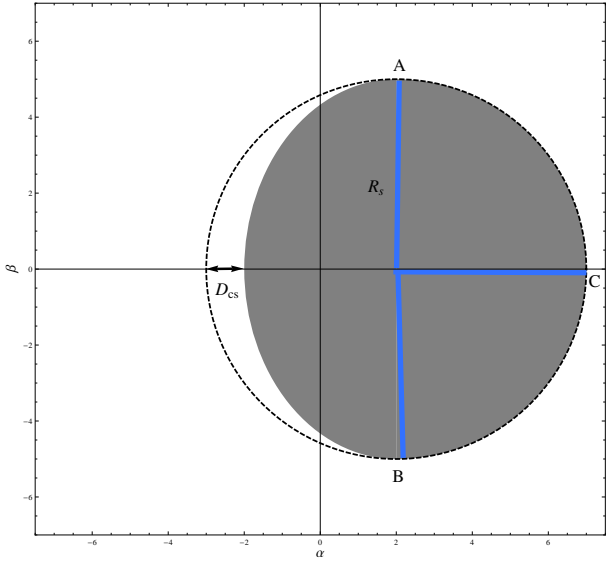}
\caption{Scheme of radius 
and distortion of the silhouette 
of rotating black hole 
with gravitomagnetic 
charge. 
The distortion parameter is defined as 
$\delta_{\rm sh}= 
D_{cs}/R_{\rm sh}$, 
where 
$R_{\rm sh}$ 
is the average radii of shadow and 
$D_{\rm cs}$ 
is the
distance of deviation 
of the shape of the 
shadow from circle 
shape. 
\label{rsdefine}}
\end{center}
\end{figure}

While considering 
the shape of the shadow 
of rotating black hole with 
non vanishing gravitomagnetic 
charge one may 
introduce the radius 
$R_{\rm sh}$ 
and the 
distortion parameter
$\delta_{\rm sh}$ 
of the silhouette related by the expression 
$\delta_{\rm sh} = D_{\rm cs}/R_{\rm sh}$. 
The schematic explanation of these parameters are shown 
in~Fig.~\ref{rsdefine}~\citep{maeda}.
%

Consider the distant observer located in the equatorial plane
of the rotating black hole black hole with non vanishing gravitomagnetic charge. Then,
\begin{eqnarray}
\label{alphapsi2} 
\alpha&=&-\xi \ ,
\\
{\rm and} 
&&\nonumber\\ 
 \label{beta3}
\beta
&=&
 \sqrt{\eta} \ .
\end{eqnarray}

\begin{figure*}[htb]
\begin{center}
\includegraphics[width=8cm]{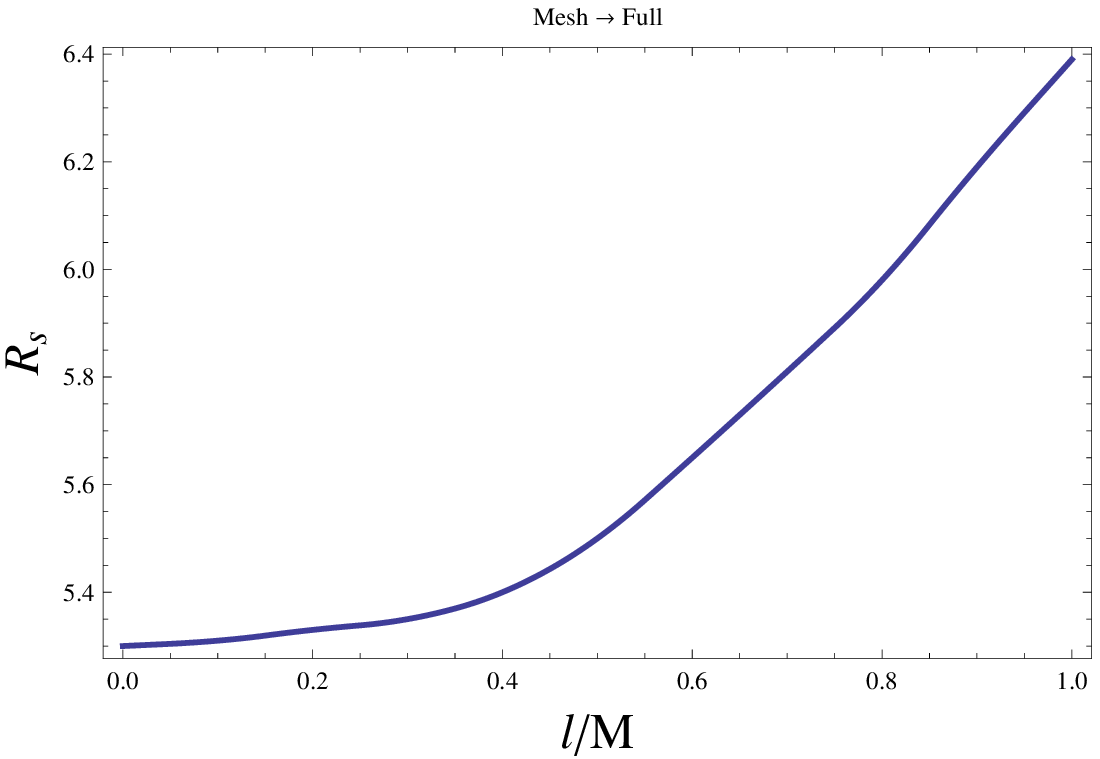}
\includegraphics[width=8cm]{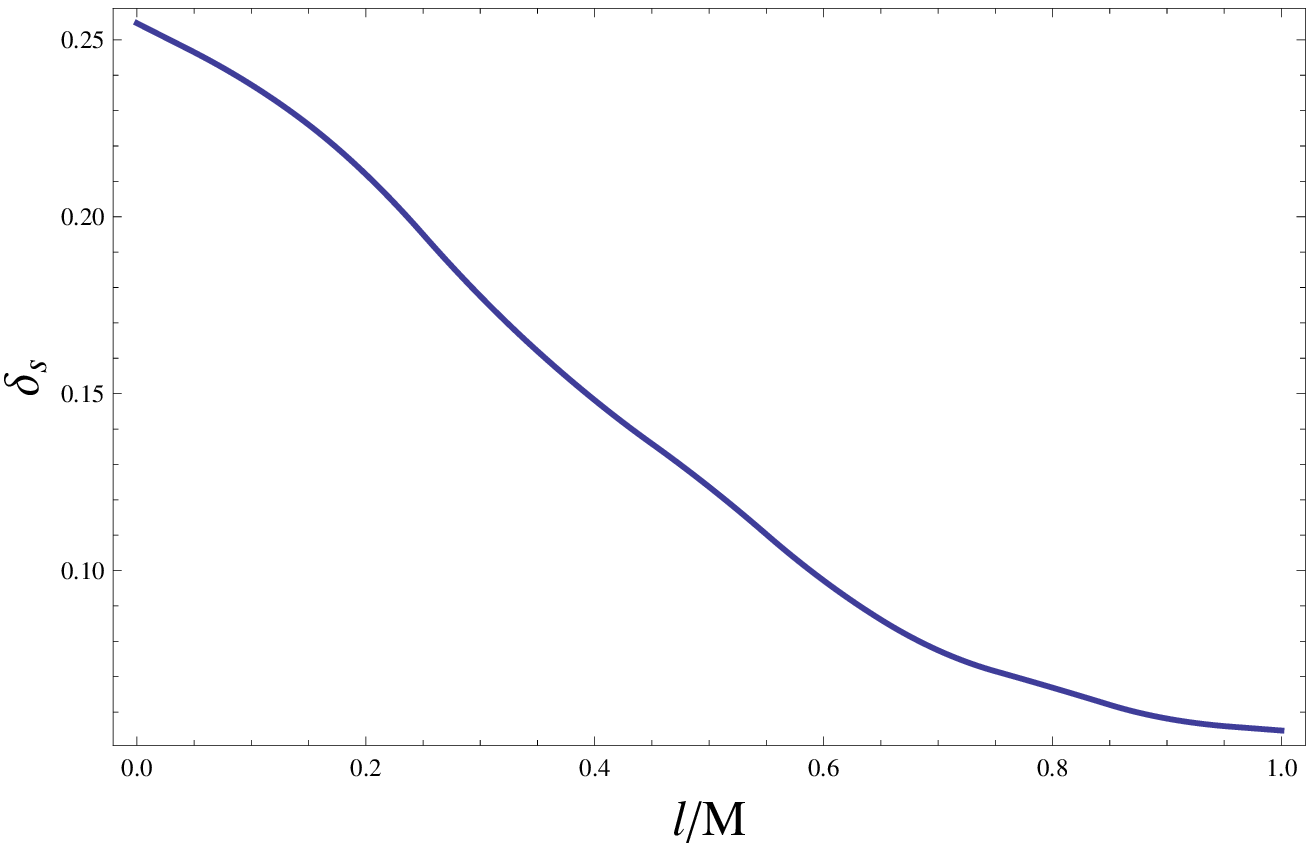}

\caption{The dependence of the average radius of the silhouette  $R_{\rm sh}$ and the distortion (deviation) parameter $\delta_{\rm sh}$ from the gravitomagnetic charge (NUT parameter). \label{rsdeltas}}
\end{center}
\end{figure*}

Obtained numerical 
results are shown 
in~Fig. \ref{aylanalar} where 
the shape of the silhouettes 
of rotating Taub-NUT black 
hole for the different values 
of the rotation parameter and 
gtavitomagnetic 
charge are presented. 
From
the plots one can see that the presence of the gravitomagnetic
charge will increase the effective size of the shadow. 
In the figure the 
shapes of the silhouette  of 
rotating black hole with the 
gravitomagnetic charge
are given for the 
different values of black hole angular momentum 
$a$:  
$a/M
= 0.5$, 
$a/M = 0.7$, 
$a/M = 0.8$, and 
$a/M = 0.99$. 
One can easily 
compare the effect of the NUT parameter 
and the black hole rotation 
parameter  
on modification of the shape of 
the shadow of black hole.
It appears they have opposite effects on black hole shadow size.
The gravitomagnetic charge increases the size of the shadow shape
while black hole's angular momentum decreases its size.
The parameters 
$R_{\rm sh}$  (radius) and 
$\delta_{\rm sh}$ (distortion) 
are shown as
functions of the gravitomagnetic charge $l$ in Fig.
\ref{rsdeltas}. 
From the dependence of 
$R_{s}$ from the NUT
parameter one can again see 
that the gravitomagnetic charge forces
to increase the size of the black hole shadow. 
The dependence of
$\delta_{s}$ from NUT charge shows 
that gravitomagnetic charge
forces to shadow to get the shape of circle than ellipse. In the
case of rotation, with the increase of black hole's angular
momentum the shape of black hole shadow takes form of ellipse
rather than circle.

\section{Particle  capture cross sections for black hole with gravitomagnetic charge}

In this section we will study the pure effect of NUT parameter
assuming that the the angular momentum of the black hole is equal
to zero. It has been shown in  the paper of~\citet{aak08} that
variables in the Hamilton-Jacobi equation for the particle motion
around NUT black hole can be separated in the equatorial plane. In
the space-time metric~(\ref{metric}) we assume that the central
object is non-rotating and particles are confined at the
equatorial plane~($a=0,\ {\rm and } \ \theta=\pi/2$).

It was first shown by~\citet{Zim89} for the spherical symmetric
case (NUT spacetime) and later in the paper of~\citet{bini03} for
the axial symmetric case (Kerr-Taub-NUT spacetime) that the orbits
of the test particles are confined to a cone with the opening
angle $\theta$ given by $\cos\theta = 2{\cal E}l/{\cal L}$. It
also follows that in this case the equations of motion on the cone
depend on $l$ only via $l^2$ \citep{bini03,aak08}.

The main point is that the small value for the upper limit for
gravitomagnetic moment has been obtained by comparing theoretical
results with experimental data as (i) $l/M\leq 10^{-24}$ from the
gravitational microlensing \citep{habibi04}, (ii) $l/M\leq
1.5\cdot10^{-18}$  from the interferometry experiments on
ultra-cold atoms \citep{ma08}, (iii) and similar limit has been
obtained from the experiments on Mach-Zehnder interferometer
\citep{kkl08} (here $M$ is the total mass of central gravitating
object).

\begin{figure}[htb]
\begin{center}
\includegraphics[width=8cm]{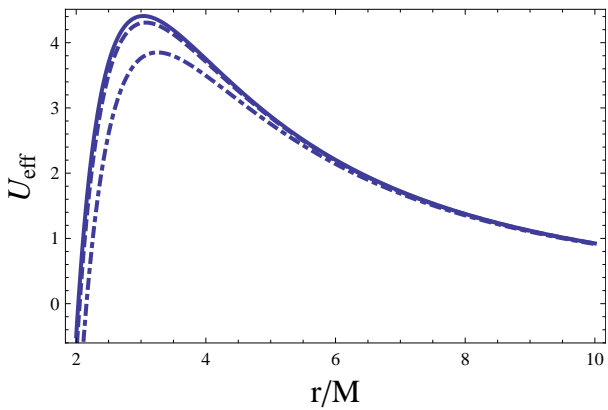}

\caption{The radial dependence of the effective potential of
radial motion of the massive particles for the different values of
the gravitomagnetic charge: solid line for $l/M=0.1$, dashed line
for $l/M=0.5$, and dot-dashed line for $l=0.9$.\label{rotrm}}
\end{center}
\end{figure}

Due to the smallness of the gravitomagnetic charge let us consider
the motion in the quasi-equatorial plane when the motion in
$\theta$ direction changes as $\theta=\pi/2+\delta\theta(t)$,
where $\delta\theta(t)$ is the term of first order in $l$, then it
is easy to expand the trigonometric functions as
$\sin\theta=1-\delta\theta^2(t)/2+{\cal O}(\delta\theta^4(t))$ and
$\cos \theta=\delta\theta(t)-{\cal O}(\delta\theta^3(t))$.
Neglecting the small terms ${\cal O}(\delta\theta^2(t))$, one can
write $\Sigma=r^2+l^2$, $\Delta=r^2-2Mr-l^2$, and $\chi=0$ and
consequently the equation of motion for the radial motion takes
the following form
\begin{eqnarray}
r^4 \left(\frac{dr}{d\tau}\right)^2 = R(r)=\left[{\cal E}^2-1-2
U_{\rm eff}(r,l,{\cal L})\right]r^4\ , \label{eqnmot}
\end{eqnarray}
where ${\cal E}$ and ${\cal L}$ is the energy and angular momentum
of the particle per unit of its mass and the quantity
\begin{eqnarray}
U_{\rm eff}(r,l,{\cal L}) = -\frac{l^2+M r }{\Sigma}+\frac{\Delta
{\cal L}^2}{2 \Sigma^2}
\end{eqnarray}
can be interpreted as effective potential of the radial motion of
the test particle at equatorial plane. The radial dependence of
the effective potential of radial motion of the massive particles
for the different values of the gravitomagnetic charge is
presented in Fig.~\ref{rotrm}.

Assuming  that the uncharged particle is moving slowly at
infinity, i.e. ${\cal E}\simeq 1$ one can easily rewrite the
expression (\ref{eqnmot}) in the following form:
\begin{eqnarray}
R(\rho)=\rho^3+\left(\tilde{l}-\tilde{\cal
L}\right)\rho^2+\tilde{\cal L} \rho+\frac{\tilde{l} \tilde{\cal
L}}{2}\ ,\label{polynom}
\end{eqnarray}
where
$$
\rho=\frac{r}{M}, \ \ \ \tilde{l}=\left(\frac{l}{M}\right)^2, \ \
\ \tilde{\cal L}=\left(\frac{{\cal L}}{M}\right)^2\ .
$$
Gravitational capture of the particle occurs for ${\cal
L}\leq{\cal L}_{\rm cr}$. For the ${\cal L}={\cal L}_{\rm cr}$
orbit spirals into a circular orbit, the radius of which is
determined by the value of the multiple root of the polynomial
(\ref{polynom}), i.e. discriminant of the later should vanish. Now
it is easy to find the expression for  ${\cal L}_{\rm cr}$ as
\begin{eqnarray}
{\cal L}_{\rm cr}^2=16M^2-6{l^2}-\frac{13 l^4}{16 M^2}\ .
\end{eqnarray}
\begin{figure}[htb]
\begin{center}
\includegraphics[width=8cm]{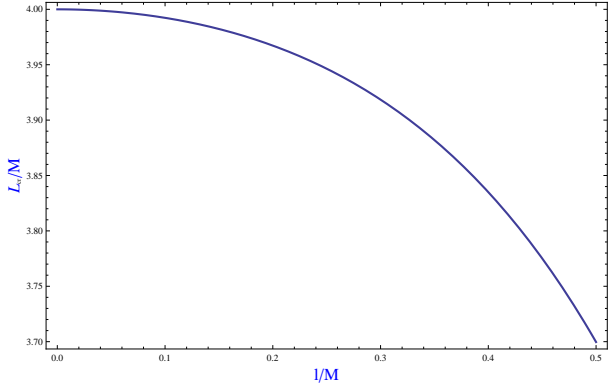}

\caption{The dependence of the critical angular momentum for
capturing by central black hole from gravitomagnetic monopole
momentum. \label{lcrit}}
\end{center}
\end{figure}

In the Fig.\ref{lcrit} the dependence of ${\cal L}_{\rm cr}$ from
dimensionless NUT parameter is presented. The dependence shows
that the presence of the gravitomagnetic charge decreases the
capture cross section for particles by black hole.

\section{Conclusion}

In this paper, we have studied the shadow of black hole with
nonvanishing gravitomagnetic charge and analyzed how the shadow of
the black hole will be  distorted by the presence of the NUT
parameter. From the numerical results we have obtained that the
NUT parameter forces to increase the size of the black hole
shadow. The dependence of the distortion parameter $\delta_{s}$
from the NUT charge shows that the gravitomagnetic charge forces
black hole's shadow to get the shape of circle than ellipse. We
have also studied the capture cross section for massive particles
by black hole with nonvanishing gravitomagnetic charge and found
its strong dependence from the NUT parameter.

\acknowledgments

AA and BA acknowledge the Akdeniz University  and TUBITAK/BIDEP
foundation for supporting local hospitality during their stay in
Antalya, Turkey. AA and BA thank the TIFR and IUCAA for the warm
hospitality during their stay in Mumbai and Pune, India. This
research is supported in part by the projects F2-FA-F113,
FE2-FA-F134 of the UzAS and by the ICTP through the OEA-PRJ-29
projects. AA and BJ acknowledge the German Academic Exchange
Service (DAAD) and the TWAS Associateship grants. This work was
supported by Akdeniz University, Scientific Research Projects
Unit.

%

\end{document}